\documentclass{ws-mpla}
\usepackage{amssymb}

\begin{document}

\markboth{Renxin Xu} {Astro-quark matter: a challenge facing
astroparticle physics}

\catchline{}{}{}{}{}

\title{Astro-quark matter: a challenge facing astroparticle physics}

\author{Renxin Xu}

\address{School of Physics, Peking University,
Beijing 100871, China; r.x.xu@pku.edu.cn}

\maketitle


\begin{abstract}

Quark matter both in terrestrial experiment and in astrophysics is
briefly reviewed.
Astrophysical quark matter could appear in the early Universe, in
compact stars, and as cosmic rays. Emphasis is put on quark star as
the nature of pulsars.
Possible astrophysical implications of experiment-discovered sQGP
are also concisely discussed.

\end{abstract}

\ccode{PACS Nos.: 21.65.Qr, 97.60.Gb, 97.60.Jd, 12.38.Mh.}

\bigskip
\rightline{\tt `` Take a stick of wood with {\it finite} extent,}%
\rightline{\tt one cuts half each day;}%
\rightline{\tt  it is expected to last out for an {\it infinitely} long time. ''}%
\rightline{\rm --- in {\em Tianxia Pian} by Zhuang Zi [Chuang Tze] ($\sim 369$BC to 286BC)}%
\bigskip

\section{Introduction: quark matter}

More than two thousand years ago, a Chinese philosopher, Zhuang
Zi, speculated that all of matter in the world are dividable.
However, at almost the same time, a Greek philosopher, Demokritos
($\sim 460$BC to 370BC), suggested that everything is made up from
elements that can not be divided any further, called {\em atoms}
(i.e., in-dividable elements).
Now, it is understood in modern physics that atoms are actually
dividable because of the detections of electron (discovered by
Thomson) and nucleus (by Rutherford), and that the dynamics in an
atom is governed by quantum mechanics.
In the standard model of particle physics, all of matter are
composed of fundamental Fermions, 6 flavors of quarks and 6
flavors of leptons, between which elementary interactions are
mediated by gauge bosons.
In spite of that, each type of these fundamental particles is
supposed to correspond to one of certain vibrational modes of
strings (typically $\sim 10^{-33}$ cm) described mathematically in
the string theory in order to explain the great discrepancy
between the quantum theory and the general relativity.
Anyway, it is well recognized that a variety of states of matter
exist in different physical conditions (e.g., Fig. 1). {\em Quark
matter} is composed of quarks (and possible gluons) as the
dominant degrees of freedom.
\begin{figure}[h]
\includegraphics*[width=.9\textwidth]{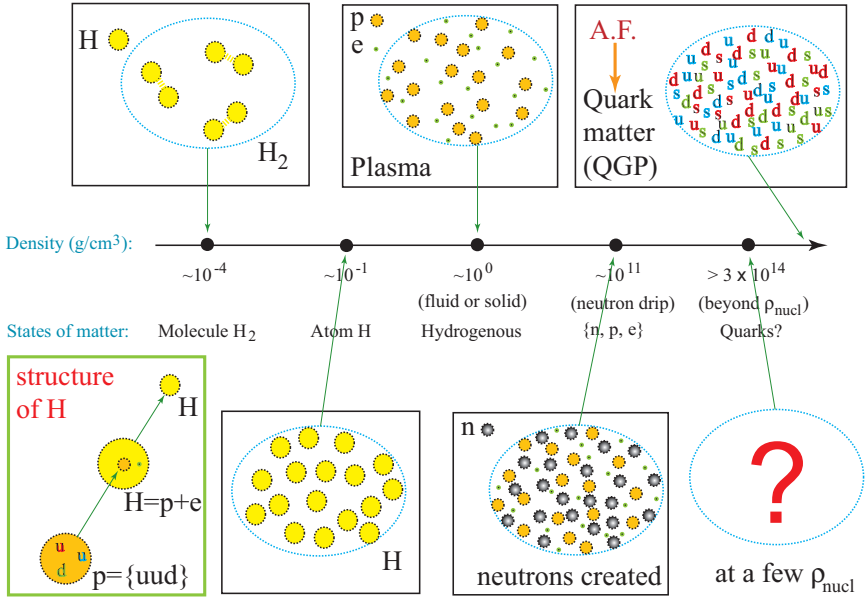}
\caption{Density effect dominated phases of matter composed simply
of electrons and protons. Temperature effect is negligible here.
It is evident that different particle degrees of freedom dominate
as the density increases. Note that one could know the
microphysics by studying the states of matter at different
densities. ``A.F.'': asymptotic freedom, ``$\rho_{\rm nucl}$'':
the nuclear density.}
\end{figure}

Quark matter could be the {\em key} to the sub-quarkian physics.
If Zhuang Zi's idea is true, we may ask naively: Could the
fundamental Fermions be dividable? What could the strings be made
of? The answers to these and other questions related (e.g., the
vacuum) would depend on the nature of quark matter, the state of
which can be tested by terrestrial experiments as well as
astrophysical observations.

Let's introduce briefly the history of quark matter.
In the standard model, quantum chromodynamics (QCD) is believed to
be the underlying theory of the elementary strong interaction,
which has two general properties. For the strong interaction in a
small scale ($\sim 0.1$ fm), i.e., in the high energy limit, the
interacting particles can be treated as being {\em asymptotically
free}; a perturbation theory of QCD (pQCD) is possible in this
case.
Whereas in a larger scale ($\sim 1$ fm), i.e., in the low energy
regime, the interaction is very strong, which might result in the
{\em color confinement}.
The pQCD is not applicable in this scale (the non-perturbative
effects are not negligible), and quarks and gluons are confined in
hadrons. Nevertheless, in this case, one can still study effectively
the color interaction: (i), the lattice formulation (LQCD), with the
discretization of space-time and on the base of QCD, provides a
non-perturbative framework to compute numerically relations between
parameters of the standard model and experimental phenomena; (ii),
QCD-based effective models with suitable Lagrangian density could
also help us to speculate the features of the low energy QCD; (iii),
phenomenological models, which rely on experimental and astrophysics
date available at low energy density, are advanced for the phase
diagram of QCD (e.g., the states of super-dense hadron and quark
matter).

The general features (asymptotic freedom and color confinement) of
QCD would result in two distinct {\em phases} of matter, depicted in
the QCD phase diagram in terms of temperature $T$ vs. baryon
chemical potential $\mu_{\rm B}$ (or baryon number density).
Hadron gas phase locates at the low energy-density limit where
both $T$ {\em and} $\mu_{\rm B}$ are relatively low, while a new
phase called {\em quark gluon plasma} (QGP) or {\em quark matter}
appears in the other limit when $T$ {\em or} $\mu_{\rm B}$ is high
although this new state of matter is still not found with
certainty yet.
It is therefore predicted that there is a kind of phase transition
from hadron gas to QGP (or reverse) at critical values of $T$ and
$\mu_{\rm B}$.
Actually a deconfinement transition is observed in numerical
simulations of LQCD for zero chemical potential $\mu_{\rm B}=0$,
when $T\rightarrow T_{\rm qcd}\simeq (150\sim 180)$ MeV.

{\em Hot quark matter studied via relativistic heavy ion
collisions}.
One way to investigate experimentally the state of quark matter is
through collision of two relativistic heavy ions. A kind of matter
with asymptotically free quarks and gluons was expected
previously, but real experimental results (e.g. RHIC, the
relativistic heavy ion collider) indicate a fireball with strong
interaction, so-called sQGP\cite{Shuryak} (strongly coupled QGP),
because of
(i) jet quenching (the suppression of one of the two jets produced
by a collision pair of energetic quarks or gluons near the edge of
the fireball implies a short mean free path of particle in the
bulk matter formed)
and (ii) elliptic flow (successful hydrodynamics computations for
the spatially anisotropic flow shows that the bulk matter could be
well approximated by perfect fluid, i.e., zero mean free path, in
case of low transverse momentum).
Besides, an interesting form of matter, the color glass condensate
(a term ``glasma'' is then coined), is argued to occur before
reaching an equilibrium state of the fireball (several typical time
scales are comparable at that time)\cite{glasma}.

The sQGP matter is surely a new state with energy densities more
than ten nuclear density and at temperatures of particle kinematic
energy of $\sim  200$ MeV.
It is worth noting that the composed ingredients in the sQGP is
quarks and gluons, which is not confined in hadrons, but the
interaction between these particles is very strong.
As argued by Csorgo\cite{Csorgo} for the Pb+Pb collisions at CERN
SPS (super proton synchrotron), this new form of matter created is
in principle not QGP theoretically predicted, but is actually quark
matter with effective degrees of freedom to be the massive (dressed)
constituent quarks instead of almost massless quarks and gluons.
The future Brookhaven RHIC and CERN LHC (the large hadron
collider) programs will certainly improve our understanding of
this kind of quark matter.

Anyway, how can we test further our theoretical view points on
sQGP by other experiments? What could be the astrophysical
implications of sQGP?

{\em Cold quark matter studied astrophysically?}
Another way to do is via observing astrophysical appearances as
well as implications of quark matter. All the possible quark
matter residual from astrophysical processes is cold though it may
be hot initially (e.g., during the early Universe).

Cold quark matter is another story.
Extremely dense quark matter could certainly be regarded as ideal
Fermi gas due to the asymptotic freedom. However, should
astrophysical quark matter at a few nuclear densities be
asymptotically dense?
In fact, it appears that the highly degenerate Fermi surface
should be unstable to form Cooper quark pairs, condensed in
momentum space, around the surface when weak attraction between
quarks is introduced. A novel state, which is similar to the
electric superconductivity, is then speculated, called color
superconductivity (CSC), for cold baryon matter at a few nuclear
densities\cite{CSC2008}.
A state of two-flavor color superconductor (2SC) may occur at
lower density, while a color-flavor locked (CFL) phase would exist
at higher density.

What if the mean free path of quarks in cold quark matter is {\em
very short} (i.e., the matter is also strongly coupled)?
This is an interesting question necessary to be raised and answered
after discovering strong interaction in the hot quark matter.

Actually, based on possible astrophysical features detected, a state
of solid quark matter was conjectured five years ago, just before
inventing the abbreviation of ``sQGP'', and the author proposed more
realistically that quark clusters (i.e., quarks are condensed in
{\em position} space rather than in momentum space) could be
essential for the {\em normal} solid state\cite{xu03}.
This idea would be natural since a short mean free path may favor
positional condensation of particles.
As illustrated in Fig. 2, in different locations of the QCD
phase-diagram, the vacuum would have different features and is
thus classified into two types: the perturbative-QCD (pQCD) vacuum
and nonperturbative-QCD (QCD) vacuum. The coupling is weak in the
former, but is strong in the latter. Quark-antiquark (and gluons)
condensations occur in QCD vacuum (i.e., the expected value of
$\langle q{\bar q} \rangle \neq 0$), but not in pQCD vacuum. The
chiral symmetry is spontaneously broken in the case that the
vacuum is changed from pQCD to QCD vacuums, and (bare) quarks
become then massive constituent ones (dressed quarks). This
theoretical points are consistent with the solid quark matter
speculation since there is no observation that the quark
deconfinement and the chiral symmetry restoration should occur
simultaneously.
\begin{figure}[h]
\includegraphics*[width=10cm]{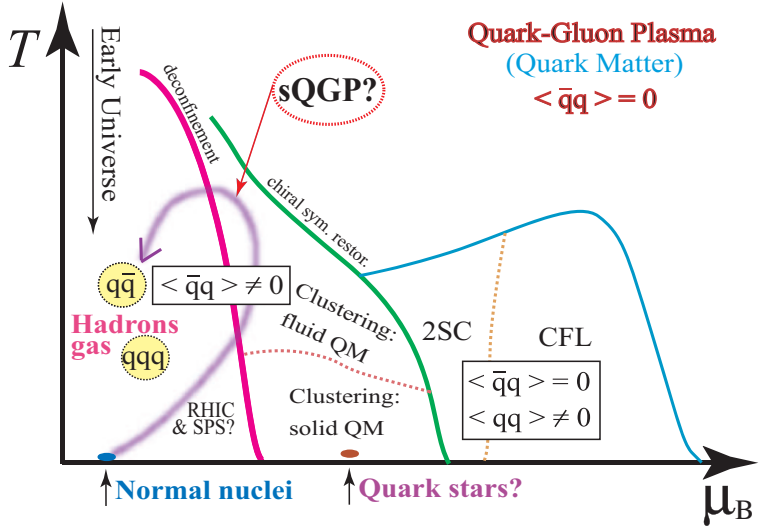}
\caption{Schematic illustration of QCD phase diagram speculated in
astrophysics.}
\end{figure}
Besides, based on the effective models, the rigidity of CSC quark
matter could not be negligible if the gap parameter modulates
periodically (i.e., the translational invariance is thus broken),
and the shear moduli of this crystalline {\em super}-solid state
could be as high as 20 to 1000 times larger than those of neutron
star crusts\cite{mrs07}. It could be interesting to observationally
distinguish between and search evidence for possible normal-solid
and super-solid states although the latter seems to be more robust
than the former from a purely theoretical point of view.

We will try an effort to review briefly the astrophysical quark
matter in the next section.
The previous short reviews\cite{xu06,xu07} could also be valuable
for some details.

\section{Quark matter in astrophysics}

There could be mostly three forms of astrophysical quark matter,
i.e., appearing in the early Universe, in compact stars, and as
cosmic rays.

\subsection{Cosmic QCD phase transition}

The symmetry in high-temperature (high-$T$) is much larger than that
in low-$T$, the vacuum could then undergo various phase-transitions
in case of symmetry breaking when the Universe cooled as it
expanded. One of the transitions relevant to astro-quark matter is a
QCD transition in the early Universe.
In the radiation-dominated universe, whose space-time is described
by Robertson-Walker metric, the temperature can be well approximated
simply by
$ T\simeq 1~{\rm MeV}/\sqrt{t}
$
, with the cosmic age $t$ in seconds.
According to LQCD simulations for case of zero chemical potential, a
quark-hadron phase-transition (QHPT, or QCD transition) took place
at temperature $T_{\rm qcd}\simeq (100\sim 200)$ MeV when the cosmic
age was $t_{\rm qcd}\simeq 10^{-5}$ s.

The cosmic QHPT is very close to an equilibrium process, since the
the relaxation time scale of color interaction, $\sim 1~{\rm
fm}/c\sim 10^{-23}$ s, is much smaller than the time interval
$t_{\rm qcd}\sim 10^{-5}$ s in which the cosmic thermodynamical
variables and expanding-dynamical curvature can change significantly
(see Ref.~\refcite{sch03} for a general review).
The key uncertainty factor is the order of QHPT, on which several
astrophysical implications depend (see next three paragraphs).
A first-order transition may proceed through bubble nucleation. The
hadronic bubbles grow, release laten energy, and could collide with
others when they are larger enough (i.e., bubbles with hadron gas
grew until they merge and filled up the whole universe in the end of
QHPT).
The horizon radius at that time is $R_{\rm h}\sim ct_{\rm qcd}\sim
10$ km. However, the typical separation between bubbles, $D_{\rm
b}$, could be much smaller than the horizon radius, that is only
$D_{\rm b}\sim 10^{-6}R_{\rm h}\sim 1$ cm according to lattice QCD
calculations where the bubble surface tension and laten heat are
included.

The cosmic QHPT may have many astrophysical consequences which would
test the physical process in turn.
Big-bang nucleosynthesis (BBN) predicts the abundances of the light
elements (D, $^3$He, $^4$He, and $^7$Li) synthesized at cosmic age
of $\sim 10^3$ s, which are observation-tested spanning {\em nine}
orders of magnitude (number ratios: from $^4$He/H$\sim 0.08$ down to
$^7$Li/H $\sim 10^{-10}$).
However, the initial physical conditions for BBN should be setted by
this QHPT. For instance, the inhomogeneities of temperature and
baryon numbers during bubble nucleation may affect the abundances
synthesized, which may clear the possible inconsistency of the light
element abundances with the CMB measurements\cite{cfo03}.
In this sense, BBN offers then a reliable probe of QHPT. As a
result, this study could provide a better determination of the
baryonic density in the universe.

The formation of quark nuggets could be another probable
consequence.
Towards the end of the QHPT, baryon-enriched quark droplets shrank,
and might remain finally to play the role of dark
matter\cite{Witten84}. Quark droplets with strangeness are
conjectured to absolutely stable, and the residual quark nuggets
could then probably be composed of strange quark matter with high
baryon density.
Can we detect such quark nuggets? This could be a meaningful project
to be done, both experimentally and theoretically, in the future.

There could be other relics of cosmic QHPT. A very interesting issue
is to study the bubble collisions which may be responsible to the
generation of gravitational waves\cite{Witten84}.
Seed magnetic fields could be produced by currents on the bubble
surface\cite{he98}.

\subsection{Quark matter and pulsars}

Although one may conventionally think that pulsars are `normal'
neutron stars, it is still an open issue whether pulsar-like stars
are neutron or quark stars\cite{madsen99,lp04,weber05}, as no
convincing work, either theoretical from first principles or
observational, has confirmed Baade-Zwicky's original idea that
supernovae produce neutron stars.

\subsubsection{Historical notes on pulsars, neutron stars, and quark stars}

Soon after the Fermi-Dirac form (in 1926) of statistical mechanics
was proposed for particles which obey Pauli's exclusion principle
(in 1925), it is Fowler (in 1926) who recognized that the electron
degeneracy pressure can balance for those stars, white dwarfs
discovered by astronomers in 1914. Only two further steps (the state
equation of a completely degenerate gas and numerically calculating
the hydrostatic equilibrium with the state equation) are needed to
pass from Fowler's discovery\cite{chandra45}, that were then carried
out by Chandrasekhar (in 1931) who found a unique mass (the mass
limit of white dwarfs).
What if the mass of a star supported by electron degenerate pressure
is greater than the Chandrasekhar limit? Landau speculated a state
of matter, the density of which ``becomes so great that atomic
nuclei come in close contact, forming one {\em gigantic nucleus}''
in 1932. A star composed dominantly of such matter is called a
``neutron'' star, and Baade and Zwicky even suggested in 1934 that
neutron stars could be born after supernovae.
A direct observational evidence, proposed by Gold in 1968, is
detecting pulsed radio beams (pulsars) due to the lighthouse effect
of spinning neutron stars, although pulsars were supposed to ``be
associated with oscillation of white dwarfs or neutron
stars\cite{psr68}''.

However, neutrons and protons are in fact {\em not} structureless
point-like particles although they were thought to be elementary
particles in 1930s; they (and other hadrons) are composed of {\em
quarks} proposed by Gell-Mann and Zweig, respectively, in 1964.
The quark model for hadrons developed effectively in 1960s, and
Ivanenko \& Kurdgelaidze\cite{ik69} began to suggest a quarkian core
in a massive neutron star. Itoh\cite{itoh70} even considered
3-flavor (\{u,d,s\}) {\em full} quark stars (now called {\em
strange} quark stars), with calculation of their hydrostatic
equilibrium.
The speculated quark stars could really exist if bulk strange quark
matter is most stable\cite{Witten84}.
Haensel, Zdunik \& Schaeffer\cite{hzs86} and Alcock, Farhi \&
Olinto\cite{afo86} then modelled strange stars, and found that these
quark stars can also have typical masses of $\sim (1-2)M_\odot$ and
radii of $\sim 10$ km, that means that {\em the pulsar-like stars
believed previously to be neutron stars might actually be quark
stars}.

How to distinguish observationally quark stars from neutron stars?
Alcock et al.\cite{afo86} thought that (1) a strange quark star
would accrete inter-stellar matter, and then a crust would form,
wrapping the strange quark star, and (2) a bare strange star can not
manifest itself as a radio pulsar because of being unable to
generate a magnetosphere.
This view was criticized by Xu \& Qiao\cite{xq98}, who addressed
that these two points are theoretically unsuitable, and that {\em
bare} strange quark stars (i.e., without crusts) are welcome for
astronomers to understand observation. A new window to distinguish
between neutron and quark stars is then opened since there are
striking differences between the exotic quark surfaces of bare quark
stars and the normal matter surfaces of neutron stars.
With regard to the possible methods to identify quark stars in
literatures, hard evidence may be obtained by noting the surface
differences since the other avenues are subject to many complex
nuclear and/or particle physics processes that are poorly known.

It is really a non-perturbative task to understand the QCD phase
diagram, to which a vast range of fundamental physics problems are
related.
The state of matter in pulsar-like stars is yet not determined.
Nevertheless, we would like to summarize various speculations about
the nature of pulsars in Fig. 3.
\begin{figure}[th]
\centerline{\psfig{file=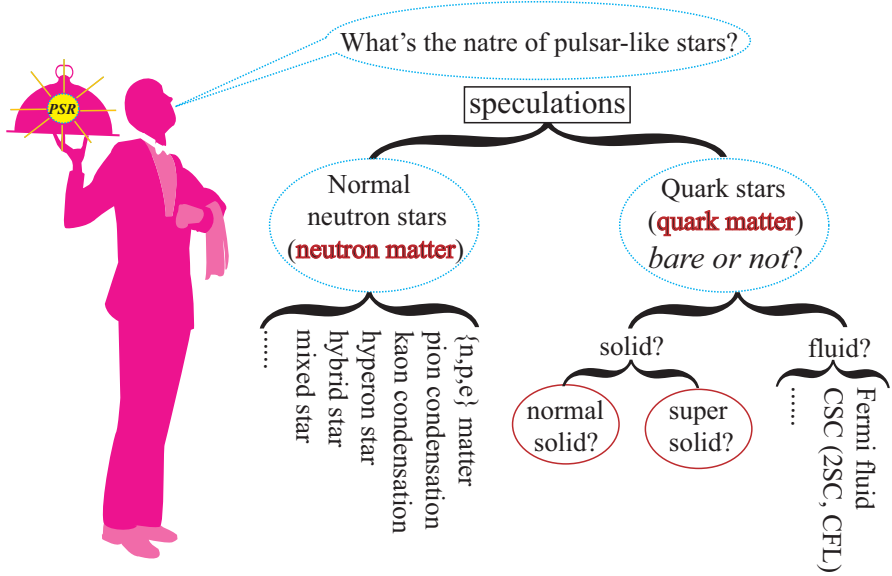,width=12cm}} \vspace*{8pt}
\caption{We are still not sure about pulsar's real internal
structure because of the uncertainty of the state of matter at
supra-nuclear densities. A variety of states have then been
speculated for the nature of pulsar-like stars, which could be
classified into two categories: neutron (or nuclear) matter
(normal neutron stars) and quark matter (quark stars).
\protect\label{fig1}}
\end{figure}

\subsubsection{Observational hints that pulsar-like stars could be quark stars}

Although quark stars seem to be `easily' ruled out from time to time
in the literatures (similar to the case of one's refraining from
smoking), we would like to address possible evidence for them since
this is still an unsolved physical and astrophysical problem mixed
with a variety of research subjects.

{\em Magnetospheric and thermal emission features: bare quark
surface?}
The RS-type vacuum gap model\cite{rs75}, with an ``user friendly''
nature, is popular and successful in explaining the radiative
behaviors of radio pulsars, which can only work in strick
conditions: strong magnetic field and low temperature on surface of
pulsars with $\Omega\cdot{\bf B}<0$. This binding energy could be
completely solved for any $\Omega\cdot{\bf B}$ if radio pulsars are
bare quark stars\cite{xq98,xqz99}.
Drifting subpulses\cite{dr99} and microstructures could be strong
evidence for RS-type sparking on polar caps, and further more, the
bi-drifting phenomena\cite{qiao04} could only be understood in a
bare quark star model.
Additionally, the bare quark surface could also help to explain a
few other observations.
Only a layer of degenerated electrons in strong magnetic fields on
bare quark surface, which can naturally reproduce non-atomic
spectra\cite{xu02} though atomic features were predicted in normal
neutron star models long before the observations. The absorption
lines of several X-ray sources (e.g., 1E1207 and SGR1806) could
originate from transition between Landau levels of
electrons\cite{xwq03}.
Besides both magnetospheric and thermal emission features, the quark
surface may help to alleviate the current difficulties in
reproducing two kinds of astronomical bursts which are challenging
today's astrophysicists to find realistic explosive mechanisms.
Because of chromatic confinement (the photon luminosity of a quark
surface is then not limited by the Eddington limit), bare quark
stars could create a lepton-dominated fireball\cite{xu05mn,ph05}
which could push the overlying matter away through photon-electron
scattering with energy as much as $\sim 10^{51}$ erg for successful
supernovae\cite{cyx07}. Asymmetric explosion in such a way may
naturally result in long-soft $\gamma$-ray bursts and in kicks on
quark stars\cite{cui07}.

{\em Mass-radius relation: low-mass quark stars?}
The striking difference between the mass-radius relations\cite{lp04}
of normal neutron stars and of (bare) quark stars is thought to be
useful for identifying quark stars, and yet it is worth paying
attention to {\em low}-mass quark stars\cite{xu05mn} since quark and
neutron stars with similar maximum masses can hardly be
distinguished observationally.
Actually, there may be some observational hints of low-mass
pulsar-like stars, which include the spin and polarization
behaviors\cite{xxw01} of PSR 1937+21, the peculiar timing
behavior\cite{xu05mn} of 1E1207, no-detection of gravitational waves
from radio pulsars\cite{xu06gw}, and small polar cap
area\cite{ycx06} of PSR B0943+10.
The detected small thermal area\cite{pavlov04} ({\em if} being
global) of central compact objects may reflect their low masses too.
Solar-mass and low-mass quark stars may form in different channels:
core-collapse explosion for the former and AIC (accretion-induced
collapse) of white dwarfs for the latter. The latter could also be
possibly the residue of cosmic QCD phase separation in the early
Universe.

{\em Evidence for solid quark matter?}
Based on a variety of observational features, a solid state of cold
quark matter was conjectured\cite{xu03}, which could be allowed in
the regime of QCD.
A star-quake occurs as strain energy develops in an evolving solid
star with rigidity, which can naturally result at least in two
observable phenomena: bursts due to energy release and swift/slow
spin jumps due to change of inertia momentum.
As noted previously\cite{xu06}, two kinds of factors could result in
the development of stress.
(i) As a quark star cools (even spinning constantly), changing state
of matter may cause a development of anisotropic pressure
distributed inside a solid matter. Such matter cannot be well
approximated by perfect fluid, and the equation governing star's
gravitational equilibrium should then not be the TOV equation. For
stars being spherically symmetric, one can introduce the difference
between radial and tangential pressures, $\Delta$. Change of
$\Delta$ would lead to no-conservation of stellar volume. Quakes in
this {\em bulk-variable} case could be very strong and may explain
the superflares of soft $\gamma$-ray repeaters\cite{xty06}.
(ii) An uniform fluid star would keep the Maclaurin figure, and the
eccentricity decreases as a star spins down. However, for a solid
star, the shear stress would prevent the star from decreasing
eccentricity during spindown, and {\em bulk-invariable} force
develops then. As demonstrated in Fig. 4, a glitch occurs too as the
stress releases.
Both bulk-invariable and bulk-variable forces could result in
decreases of moment of inertia, and therefore in pulsar glitches.
These two kinds of forces could trigger normal glitches\cite{z04} if
they are relatively stronger than a critical stress, but might only
conduce to slow glitches\cite{px07} if weaker.
\begin{figure}[h]
\includegraphics*[width=.9\textwidth]{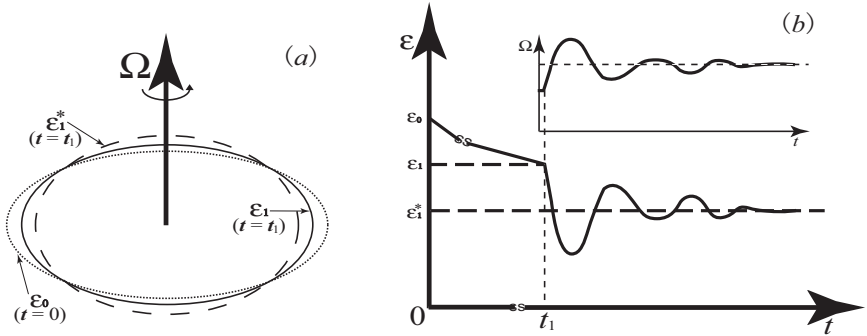}
\caption{An illustration of bulk-invariable force induced quake of a
solid quark star. The Maclaurin figure determines the eccentricities
of $\varepsilon_0$ and $\varepsilon_1^*$. A star-quake with real
eccentricity $\varepsilon_1$ occurs at time $t_1$. The
$\varepsilon$-oscillation damps quickly due to high
effective-viscosity during a real quake process.}
\end{figure}

{\em Others.}
(i). The mass of quark star could be low due to self-confinement,
but also be high due to the high incompressibility of solid quark
matter. Stellar models for {\em non-perfect} fluid matter would then
be necessary to explain pulsar's mass $\gtrsim 2M_\odot$ (e.g., the
mass\cite{nice2005} of PSR J0751+1807).
(ii). Solid quark stars might sustain maximum dimensionless
quadrupoles up to a few times $10^{-4}$, being much higher than that
of neutron stars\cite{Owen05}, and their gravitational waves are
then stronger. However, the gravitational radiation from low-mass
solid quark stars, even with maximum quadrupoles, should be
negligible\cite{xu06gw}, and no such a signal could thus be found in
the data from the fourth LIGO science run\cite{ligo08}.
(iii). The SGR-like superflares in distant galaxies could be the
dominated mechanism for short-hard $\gamma$-ray bursts since the
released energy via such giant quakes could be as high as $\sim
10^{48}$ ergs\cite{xty06}.

\subsubsection{To identify a quark star in the future?}

Four ways to test quark star model were recorded in an ``arXiv.org''
paper (astro-ph/0410652), which was published two years
later\cite{xu06asr}.
I will follow these points because I think they are {\em not} behind
the times.
(i). Dust emission around pulsar-like stars.
Disks around quark stars are suggested although the details are
still not yet clear. Sub-millimeter emission at temperature $\sim
0.1$ eV could be expected\cite{xu06asr}, and the Spitzer
detection\cite{wcd2006} could be related.
(ii). Determination the radii of distant pulsar-like stars.
We can hardly conclude a detection of pulsar-like star with radius
$\lesssim 5$ km through bolometric observations because of the
debate: are the detected thermal components global or local?
Nonetheless, future X-ray interference telescopes (e.g., {\em
MAXIM}) may make it possible to constrain real stellar radii.
(iii) Gravitational wave detection.
Gravitational wave emission associated with the energetic
superflares of SGRs may be gathered from the LIGO
data\cite{Horvath05}, and an upper limit\cite{ligo07} of $7.7\times
10^{46}$ erg for the 92.5 Hz QPO of SGR 1806-20 had been put.
However, significant high-frequency ($\sim 10^{(3-7)}$ Hz)
gravitational waves during the initial spikes of superflares may be
radiated since most of the photon emission is also radiated
initially ($\lesssim 1$ ms).
(iv). Searching for sub-millisecond pulsars.
Normal neutron stars can {\rm not} spin at period $\lesssim 0.5$ ms,
but low-mass bare quark stars can\cite{xu05mn}, even $\lesssim 0.1$
ms.
It is then a very clear evidence for quark stars if we could detect
sub-millisecond pulsars by advanced radio telescope (e.g., the
``FAST'') in the future.

Quark stars with companions of white dwarfs or quark planets are
expected\cite{xu05mn,xu06gw}. A planet around a white dwarf could be
a quark planet if no thermal emission predicted for normal planets
is detected.
Precise pulsar timing and advanced IR/sub-millimeter detecting are
then necessary to test these ideas.

\subsubsection{Quark stars v.s. neutron stars}

At this stage of study, most the observations could be principally
understood in both neutron and quark star models. However, there are
some features which have not been extensively studied. A comparison
between neutron and quark star models is summarized in Table 1. It
seems that the neutron star model is more questionable.
\begin{table}[h]
\tbl{Neutron stars vs. Quark stars: to explain the observational
features of pulsar-like stars in these two kinds of models.}
{\begin{tabular}{@{}llccc@{}} \toprule & Phenomena & Normal & (solid) & Note \\
& observed & neutron stars & quark stars &  \\
\colrule
Radio pulsars: & magnetospheric emission & ok? & ok? & e$^\pm$ plasma \\
 & normal glitch & vortex (un)pinning & star-quake & to be tested \\
 & slow glitch & ??? & in low-mass quark star & not in NS model \\
 & (bi)-drifting sub-pulses & binding?? & binding! & surface condition \\
 & (free) precession & damped? & no damping & rigid or not \\
 & timing noise & high in msPSRs? & solar or low mass & random torque \\
AXPs/SGRs$^*$: & energy source & B-field & gravity \& strain & magnetar? \\
 & burst with glitch $10^{-6}$ & ? & AISq$^*$ & sometimes \\
 & super-flare & high-B magnetar? & giant-quake? & \\
CCOs$^*$: & age discrepancy & ? & quark star with fossil disk & \\
 & erratic timing & ? & torque by disk & \\
DTNs$^*$: & non-atomic feature & high $B$ or $Z$? & bare quark stars! & \\
{\em Thermal radii} & why small? & polar cap? & low-mass quark stars & local or global \\
APXPs$^*$: & ADmsPSRs$^*$ & ok? & low-mass quark star? & spin up \& down \\
XRBs$^*$: & bursts & nuclear power & crusted quark star? & \\
Sub-msPSR$^*$: & spuper-Kepler spin & no! & possible & prediction (QS) \\
Others: & supernova & $\nu$-driven?? & $\gamma$-driven? & not successful \\
 & MACHOs$^*$ & ? & (low-mass) quark stars? & \\
 & UHECRs$^*$ & ? & strangelets? & \\
\botrule
\end{tabular}}
$^*$AXPs/SGRs: anomalous X-ray pulsars/soft $\gamma$-ray repeaters; %
CCOs: compact central objects; %
DTNs: dim thermal ``neutron stars''; %
APXPs: accretion-powered X-ray pulsars; %
XRBs: X-ray bursters; %
Sub-msPSRs: sub-millisecond pulsars; %
MACHOs: massive compact halo objects; %
UHECRSs: ultra-high energy cosmic rays; %
AISq: accretion-induced star-quake.
\end{table}

Let's choose one, discussed previously\cite{xu06asr}, in Table 1.
The timing noise is strongest in AXPs/SGRs (slowest rotators), but
is weakest in millisecond pulsars (slowest rotators); the noise
level of normal pulsars (moderate rotators) is in between those two.
This is strange in the neutron star model since the torque
variability increases with Reynolds numbers (and thus spin
frequency)\cite{mp2007}.
While, this observation could be nature in the quark star model: the
mass of most of millisecond pulsars could be $\ll 1M_\odot$, and
there should be debris disks around AXPs/SGRs.

\subsection{Quark nuggets in cosmic rays}

Two scenarios of quark nuggets in cosmic rays are discussed in the
literatures:
(i) to overcome the difficulty\cite{ml03,xw03} beyond the GZK cutoff
of the ultra-high energy cosmic rays with energy $>10^{19}$ eV, and
(ii) several exotic cosmic ray events reported by balloon and
mountain experiments to be possible candidates of strangelets (the
doubly charged event\cite{ams03} detected by the AMS experiment in
space could be a special one).
Quark nuggets could be produced in the cosmic QCD transition, might
originate from the collisions of two quark stars, and could also be
ejected by supernova explosions.
It is worth noting that the massive compact halo objects (MACHOs)
discovered through gravitational microlensing\cite{alcock93} could
probably also be low-mass quark stars formed by evolved stars,
rather than quark nuggets born during the QHPT\cite{ban03}, if
pulsar-like stars are actually quark stars.

\subsection{Others}

(i). Quark matter residues from the cosmic QCD separation could be
candidate of cold dark matter\cite{Witten84}.
(ii). Quark molecular dynamics (qMD) develops to test the
phenomenological interaction-models for hot sQGP. An extrapolation
of this to studying cold quark matter may result in a solid
state\cite{xu07}.
(iii). What if a quark nugget collides with or goes through the
Earth? It is interesting to see the seismic\cite{hrt2006} or other
responses of such compact object from space.

\section{Conclusions}

The idea of quark matter has great implications in astrophysics.
Both perturbative and nonperturbative QCD studies would be involved
in understanding the QCD phase diagram, and the hot sQGP discovered
in RHIC could surely help to understand cold quark matter in
astrophysics.

It is a pity that the real state of matter in pulsar-like stars is
still not determined confidently because of the uncertainty about
cold matter at supranuclear density, even 40 years after the
discovery of pulsar. Nuclear matter (related to {\em neutron stars})
is one of the speculations for the inner constitution of pulsars
even from the Landau's time more than 70 years ago, but quark matter
(related to {\em quark stars}) is an alternative due to the fact of
asymptotic freedom of interaction between quarks as the standard
model of particle physics develops since 1960s. Possible
observational evidence/hints that pulsar-like stars could be quark
stars are summarized, with the inclusion of achievable clear
evidence for quark stars in the future.
A solid state of cold quark matter is emphasized, and I focused on
the work of my group and feel sorry for neglecting many interesting
references due to the page limit.

There are actually three ways to study QCD phases: lattice QCD
simulations, effective QCD models, and phenomenological models.
A combined study of these three should be {\em very} necessary to
know the real QCD diagram.

\section*{Acknowledgments}
I acknowledge the contributions by my colleagues at the pulsar group
of PKU. The work is supported by NSFC (10573002, 10778611), the Key
Grant Project of Chinese Ministry of Education (305001), and by LCWR
(LHXZ200602).


\begin{thebibliography}{0}

\bibitem{Shuryak}
E. V. Shuryak, in {\it Proceedings of Continuous Advances in QCD},
hep-ph/0608177.

\bibitem{glasma} L. McLerran, {\it Nucl. Phys.} {\bf A787}, 1c (2007).

\bibitem{Csorgo} T. Csorgo, {\it Nucl. Phys. Proc. Suppl.} {\bf 92},
62 (2001).

\bibitem{CSC2008} M. G. Alford, K. Rajagopal, T. Schaefer, A. Schmitt,
{\it Rev. Mod. Phys.} in press (arXiv:0709.4635) (2008).

\bibitem{xu03}
R. X. Xu, {\it ApJ} {\bf 596}, L59 (2003).

\bibitem{mrs07}
M. Mannarelli, K. Rajagopal, R. Sharma, {\it Phys. Rev.} {\bf D76},
4026 (2007).

\bibitem{xu06}
R. X. Xu, {\it Chin. J. A\&A Suppl.} {\bf 6}, 279 (2006).

\bibitem{xu07}
R. X. Xu, {\it Adv. Space Res.} {\bf 40}, 1453 (2007).

\bibitem{sch03}
D. J. Schwarz, {\it Annalen der Physik} {\bf 12}, 220 (2003).

\bibitem{cfo03}
R. H. Cyburt, B.~D. Fields, K.~A. Olive, {\it Phys. Lett.} {\bf
B567}, 227 (2003).

\bibitem{Witten84}
E. Witten, {\it Phys. Rev.} {\bf D30}, 272 (1984).

\bibitem{he98}
M. Hindmarsh, A. Everett, {\it Phys. Rev.} {\bf D58}, 3505 (1998).

\bibitem{madsen99}
J. Madsen, in {\it Hadrons in Dense Matter and Hadrosynthesis}, p.
162 (Springer, 1999).

\bibitem{lp04}
J. M. Lattimer, M. Prakash, {\it Science} {\bf 304}, 536 (2004).

\bibitem{weber05}
F. Weber, {\it Prog. Part. Nucl. Phys.} {\bf 54}, 193 (2005).

\bibitem{chandra45}
S. Chandrasekhar, {\it ApJ.} {\bf 101}, 1 (1945).

\bibitem{psr68}
A. Hewish, S. J. Bell, J. D. Pilkington, P. F. Scott, R. A. Collins,
{\it Nature} {\bf 217}, 709 (1968).

\bibitem{ik69}
D. Ivanenko, D. F. Kurdgelaidze, {\it Lett. Nuovo Cimento} {\bf 2},
13 (1969).

\bibitem{itoh70}
N. Itoh, {\it Prog. Theor. Phys.} {\bf 44}, 291 (1970).

\bibitem{hzs86}
P. Haensel, J. L. Zdunik, R. Schaeffer, {\it A\&A} {\bf 160}, 121
(1986).

\bibitem{afo86}
C. Alcock, E. Farhi, A. Olinto, {\it ApJ.} {\bf 310}, 261 (1986).

\bibitem{xq98}
R. X. Xu, G. J. Qiao, {\it Chin. Phys. Lett.} {\bf 15}, 934 (1998).

\bibitem{rs75}
M. A. Ruderman, P. G. Sutherland, {\it ApJ.} {\bf 196}, 51 (1975).

\bibitem{xqz99}
R. X. Xu, G. J. Qiao, B. Zhang, {\it ApJ.} {\bf 522}, L109 (1999).

\bibitem{dr99}
A. Deshpande, J. Rankin, {\it ApJ.} {\bf 524}, 1008 (1999).

\bibitem{qiao04}
G. J. Qiao, K. J. Lee, B. Zhang, R. X. Xu, H. G. Wang, {\it ApJ.}
{\bf 616}, L127 (2004).

\bibitem{xu02}
R. X. Xu, {\it ApJ.} {\bf 570}, L65 (2002).

\bibitem{xwq03}
R. X. Xu, H. G. Wang, G. J. Qiao, {\it Chin. Phys. Lett.} {\bf 20},
314 (2003).

\bibitem{xu05mn}
R. X. Xu,, {\it MNRAS} {\bf 356}, 359 (2005).

\bibitem{ph05}
B. Paczy\'nski, P. Haensel, {\it MNRAS} {\bf 362}, L4 (2005).

\bibitem{cyx07}
A. B. Chen, T. H. Yu, R. X. Xu, {ApJ.} {\bf 668}, L55 (2007).

\bibitem{cui07}
X. H. Cui, H. G. Wang, R. X. Xu, G. J. Qiao, {\it A\&A} {\bf 472}, 1
(2007).

\bibitem{xxw01}
R. X. Xu, X. B. Xu, X. J. Wu, {\it Chin. Phys. Lett.} {\bf 18}, 837
(2001).

\bibitem{xu06gw}
R. X. Xu, {\it Astropart. Phys.} {\bf 25}, 212 (2006).

\bibitem{ycx06}
Y. L. Yue, X. H. Cui, R. X. Xu, {\it ApJ.} {\bf 649}, L95 (2006).

\bibitem{pavlov04}
G. G. Pavlov, D. Sanwal, M. A. Teter, in {\it Young Neutron Stars
and Their Environments}, IAU Symposium no. 218, p.239 (2004).

\bibitem{xty06}
R. X. Xu, D. J. Tao, Y. Yang, {\it MNRAS} {\bf 373}, L85 (2006).

\bibitem{z04}
A. Z. Zhou, R. X. Xu, X. J. Wu, N. Wang, {\it Astropart. Phys.} {\bf
22}, 73 (2004).

\bibitem{px07}
C. Peng, R. X. Xu, {\it MNRAS} in press (arXiv:0708.2482), (2008).

\bibitem{nice2005}
D. Nice, et al., {\it ApJ}. {\bf 634} 1242 (2005).

\bibitem{Owen05}
B. J. Owen, {\it Phys. Rev. Lett.} {\bf 95}, 211101 (2005).

\bibitem{ligo08}
B. Abbott, et al., {\it Phys. Rev.} {\bf D77} 2001 (2008).

\bibitem{xu06asr}
R. X. Xu, {\it Adv Spac. Res.} {\bf 37}, 1992 (2006).

\bibitem{wcd2006}
Z. X. Wang, D. Chakrabarty, D. L. Kaplan, {\it Natur} {\bf 440}, 772
(2006).

\bibitem{Horvath05}
J. E. Horvath, {\it Modern Physics Letters} {\bf A20}, 2799 (2005).

\bibitem{ligo07}
B. Abbott, et al., {\it Phys. Rev.} {\bf D76} 2003 (2007).

\bibitem{mp2007}
A. Melatos, C. Peralta, {\it ApJ.} {\bf 662}, 99 (2007).

\bibitem{ml03}
J. Madsen, J. M. Larsen {\it Phys. Rev. Lett.} {\bf 90} 121102
(2003).

\bibitem{xw03}
R. X. Xu, F. Wu {\it Chin. Phys. Lett.} {\bf 20} 806 (2003).

\bibitem{ams03}
V. Choutko (AMS01 Collaboration) in {\it Proc. 28th Int. Cosmic Ray
Conf.}, eds.~ T. Kojita {\it et al.} p. 1765 (IUPAP, 2003).

\bibitem{alcock93}
C. Alcock, et al., {\it Nature} {\bf 365}, 621 (1993).

\bibitem{ban03}
S. Banerjee, et al., {\it MNRAS}, {\bf 340}, 284 (2003).

\bibitem{hrt2006}
E. T. Herrin, D. C. Rosenbaum, V. L. Teplitz, {\it Phys. Rev.} {\bf
D73}, 043511 (2006).

\end{thebibliography}
\end{document}